\documentclass[proof]{WileyASNA-v1}

\articletype{Article Type}%

\received{to be inserted}
\revised{to be inserted}
\accepted{to be inserted}

\newcommand\E{1E\thinspace1740.7$-$2942}%

\begin{document}

\title{On the behavior of the black hole candidate \E's corona based on long-term INTEGRAL database}

\author[1]{Paulo Eduardo Stecchini*}

\author[2]{Jurandi Le\~ao$^+$}

\author[3]{Manuel Castro}

\author[4]{Flavio D'Amico}

\authormark{Stecchini, P. E. \textsc{et al}}

\address[1,2,4]{
                \orgname{Divis\~ao de Astrof{\'{\i}}sica/INPE},
                \orgaddress{
                            \state{~~~~~~~~~~~~~~~~~~~~~~~Avenida dos Astronautas 1758, 12227-010, S\~ao Jos\'e dos Campos-SP},
                            \country{Brazil}
                           }
                }

\address[3]{
            \orgname{Instituto de Computa{\c{c}}{\~a}o/UNICAMP},
            \orgaddress{
                        \state{Cidade Universit\'aria Zeferino Vaz, 13083-852, Campinas-SP},
                        \country{Brazil}
                       }
           }

\corres{*\email{paulo.stecchini@inpe.br}}

\presentaddress{+IFSP, Av. Bahia 1739, 11685-071 Caraguatatuba-SP, Brazil}

\abstract
         {        
          One of the most straightforward ways to explain the hard X-ray spectra
          observed in X-ray binaries is to assume that comptonization of soft photons
          from the accretion disk is occurring. The region where this
          process takes place, called the corona, is
          characterized by only two parameters: its thermal energy $kT$ and its
          optical depth $\tau$.  
          Hard X-ray spectra analysis is, thus, an imperative tool in diagnosing the behavior of these parameters. 
          The lack of consistency in obtaining/analysing long-term databases, however, may have been hindering this kind of characterization from being attained.
          With the aim of better understanding the corona behavior in the black hole
          candidate \E, we performed a homogeneous analysis for a
          large hard X-ray data set from the ISGRI telescope on-board the INTEGRAL satellite.
          Results from modelling the spectra show that, for most of our sample, unsaturated thermal
          comptonization is the main mechanism responsible for the hard
          X-ray spectra observed in \E. Moreover, such extensive database allowed us to produce what is probably the longest hard X-ray light curve of \E\space and whose units -- due to recent findings regarding dynamical quantities of the system -- could be expressed in \% of Eddington's luminosity.
         }   


\maketitle

\section{Introduction}

The object \E\space (or simply 1E) is one amongst many strong hard X-ray emitting sources in the Galactic Center  region. Although practically none is known about its probable counterpart, 1E is believed to be a black hole in a X-ray binary system due to its spectral similarities to Cyg\,X-1. Previous studies with relatively large data sets show that 1E spends most of its time in the so-called low/hard state of emission (e.g. \citealp{santo2005}, \citealp{manu2014}, \citealp{dudu2017}), a state in which the hard spectra are well described by comptonization models.

In this present study, we report on the analysis of a large hard X-ray database of 1E conducted to better comprehend how the physical quantities that characterize the corona region of the system behave. Given the spectral coverage and the amount of spectra that such analysis demands, we made use of public available data from the ISGRI telescope \citep{lebrun2003}, which, due to the regular monitoring of the source by the INTEGRAL {\citep{winkler2003}} satellite, certainly provides the largest hard X-ray spectra database of 1E. The IBIS software \citep{goldwurm2003} was used to homogeneously reduce a set of 479 INTEGRAL revolutions, that spans approximately 15 years; the fitting procedures were then performed within the analysis package environment XSPEC \citep{arnaud1996}. By applying simple and widely used models to fit the data, we were able to verify that unsaturated comptonization describes most hard X-ray spectra of 1E very well, for a wide variety of spectral indices or luminosities. Other results, as well as details on data selection and analysis, are described in the following sections.

\section{Data selection and analysis}
We retrieved from the INTEGRAL database all observations (or revolutions) comprised between $\sim$\,2003--2017 that contained \E\space in the field-of-view. With the aid of ISGRI tools for data reduction, an automatic script was developed to reduce (i.e. extract workable 20--200\,keV spectra from) these revolutions; a total number of 479 spectra of 1E were obtained. From these, 392 remained after we excluded those whose count's signal-to-noise ratio in the 20--200\,keV band were below a threshold of 5. Another script was developed, under the XSPEC suite, to automatically perform the fitting to the 392 spectra by two models: a phenomenological power-law (\texttt{powerlaw} in XSPEC notation) and a model that deals with comptonization of low energy X-ray seed photons by an electron corona, \texttt{comptt} \citep{Tita94}. The first is parametrized by the spectral index, $\Gamma$, and the second is dependent on the thermal energy $kT$ and optical depth $\tau$ of the corona, as well as on the accretion disk's inner temperature. This latter parameter was let free to vary only between 0.1 and 0.4 keV based on previous reported values (e.g. \citealp{manu2014}). Considering acceptable fits those whose $\chi^2_{\text{red}}$\,$\leq$\,2, 283 spectra were fitted by \texttt{powerlaw}, 284 by \texttt{comptt} and 250 by both models jointly. 

\begin{figure}[ht]
\centerline{\includegraphics[width=0.9\columnwidth,keepaspectratio,angle=0]{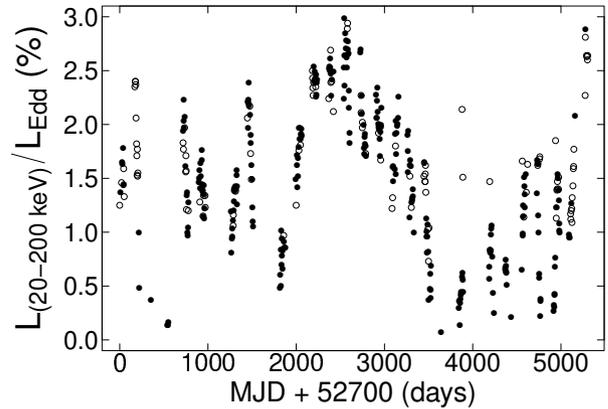}}
\caption{
         A light curve in units of \% of Eddington's luminosity for \E\space built
         from our database. The curve displays 392 points out of the
         479 initial observations. For the filled points, fluxes were calculated after a \texttt{powerlaw} fit. For the unfilled points (109 out of 392), single power-law adjustments did not
        return acceptable fit qualities and a more laborious process was used to derive the correct
         fluxes.
        } 
\label{Figure01}
        \end{figure}

Before proceeding to the comptonization analysis, we have computed the 20--200\,keV flux for each of the spectra fitted by the \texttt{powerlaw} in order to build a light curve. Given that estimates of the distance to the system (8\,kpc, \citealp{tetarenko2020}) and of the black hole mass (5\,M$_{\odot}$, \citealp{dudu2020}) have been recently reported, we are able to present this light curve in units of $L$/$L_{\text{Edd}}$. For the 109 spectra not fitted -- within the $\chi^2_{\text{red}}$ criterion mentioned -- by a single power-law, polynomial adjusts were made so the fluxes could be calculated and a more complete light curve could be produced. The resulting curve, with 392 points covering from $\sim$MJD\,52700 to $\sim$MJD\,58000, is shown in Figure \ref{Figure01}.

\section{Comptonization and the Corona in \E}

Comptonization is one major radiative process in astrophysics \citep{RybLi1979}. The implementation of the simplest comptonization model in XSPEC came very quickly ({\tt compst}; \citealp{SunyTita80}), and was promptly used to explain the high energy spectrum of Cyg\,X-1, a very well studied black hole. An improvement to this model, \texttt{comptt}, demanded some more time for implementation ($\sim$\,1994) and still is, despite being over 25 years old now, one of the most used comptonization models.

Our goal is to study the corona in \E\space by correlating the physical quantities provided by  \texttt{comptt} with the spectral indices obtained from the \texttt{powerlaw} fit. The basic approach is explained as follows.

If an up-scatter (i.e., a photon gains energy from an electron) occurs in a (so-called) thermal regime by a non-relativistic population of electrons, \citet{TitaLyu95} show that the slope -- i.e. the power-law index -- of the resulting spectrum shape must be tied to the values of $kT$ and $\tau$ from the corona. An important quantity, that may be used to discriminate the comptonization's regime and to correlate the physical parameters of the corona with the spectral index observed, is the Compton parameter $y$. This parameter is (roughly) defined as the average change in a photon's energy multiplied by the average number of scatterings a photon will suffer before leaving the corona;  a general expression for computing $y$ from $kT$ and $\tau$  is found, e.g., in Equation 2 of \citet{Petrucci2001}.  In an unsaturated comptonization regime, $y$ is likely to assume values of $\approx$\,1 and  the expected spectral index may be calculated from it, as shown in the equation for $\Gamma_1$ presented in pp. 1281 of \citet{Steiner09}. To explore the comptonization nature of our spectra, we computed $y$ from each $kT$--$\tau$ pair obtained from the \texttt{comptt} fit, calculated the ``predicted'' indices (to be called the comptonization $\Gamma$ or $\Gamma_{\text{comp}}$) and further compared these latter with the indices provided by the \texttt{powerlaw} fits. For this comparison to be possible, instead of using the 284 spectra fitted by only \texttt{comptt}, we proceeded with the 250 spectra fitted by both models.

One additional criterion was employed prior to using the \texttt{comptt} output parameters: it is pointed out in \citet{HuaTita95} and \citet{TitaLyu95} that, to assure that only acceptable physical values are being provided by the model, the pairs $kT$--$\tau$ must satisfy certain relations. In other words, there is an applicability region beyond which \texttt{comptt} results should not be used (a diagram of this region can be found in Figure 7 of \citealp{HuaTita95}). When placing the 250 spectra in this diagram (that we will refer to soon), 171 were within the zone of applicability and, thus, continued in the analysis. 

\begin{figure}[ht]
\centerline{\includegraphics[width=0.9\columnwidth,keepaspectratio,angle=0]{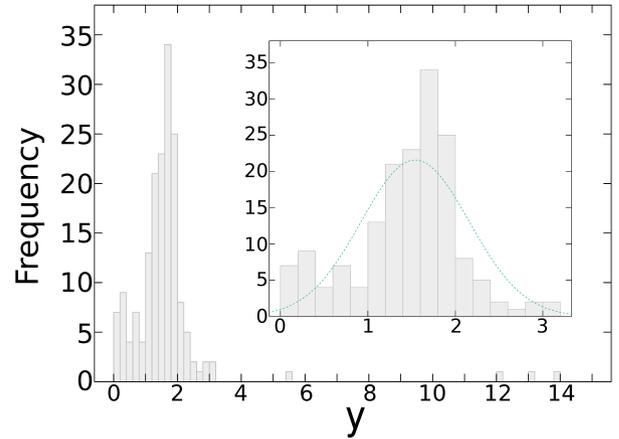}}
\caption{
         Histogram of $y$ for 171 spectra within the applicability region of \texttt{comptt}. A zoom for 0\,$\lesssim$\,$y$\,$\lesssim$\,3 is shown, as well as a normal distribution (in green) built from the median and deviation of the values in this specific range.
         \label{Figure02}
        } 
\end{figure}

\begin{figure}[ht]
\centerline{\includegraphics[width=0.9\columnwidth,keepaspectratio,angle=0]{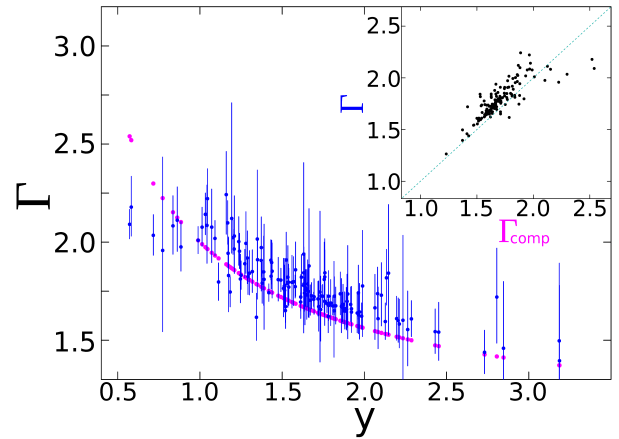}}
\caption{
         Compton parameter $y$ against the calculated $\Gamma_{\text{comp}}$ (magenta) and $\Gamma$ provided from the fits (blue).
         The inset shows, following the same colours, $\Gamma$ plotted against $\Gamma_{\text{comp}}$. For reference on their correlation,  a straight line that passes through the origin and has slope 1 is indicated (in green). 
         \label{Figure03}
        } 
\end{figure}

\begin{figure}[ht]
\centerline{\includegraphics[width=0.9\columnwidth,keepaspectratio,angle=0]{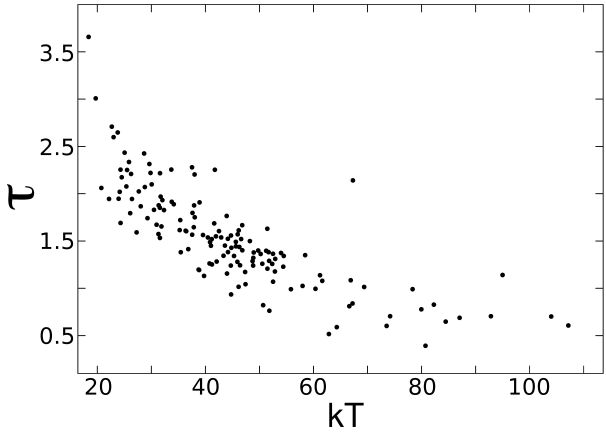}}
\caption{
         Values of the corona's thermal energy $kT$ and optical depth $\tau$, provided by the \texttt{comptt} model for the 144 selected spectra.
         \label{Figure04}
        } 
\end{figure}

The compton parameter $y$ frequency distribution for the remaining 171 spectra is presented in Figure \ref{Figure02}. A zoom, excluding the sparse occurrences above $y$\,$\approx$\,3, is also shown. The median value (and dispersion) for this particular range is 1.54\,$\pm$\,0.59. These values already indicate that unsaturated comptonization is dominant amongst the spectra. However, since establishing which values of $y$ are to be considered close to 1 is a bit vague, we proceeded and calculated $\Gamma_{\text{comp}}$ for all 171 data points. Once this expression should only be valid for cases in which $y$\,$\sim$\,1, by using $\Gamma$ from the \texttt{powerlaw} fit as a proxy and comparing $\Gamma_{\text{comp}}$ with $\Gamma$, we can constrain better what range of $y$ values represent the unsaturated regime for 1E. Hence, we select only those whose two $\Gamma$ values are coincident within 20\% of agreement. The result of this procedure, which selected 144 spectra, is shown in Figure \ref{Figure03}. The median and dispersion values for all the quantities displayed are: $y$\,=\,1.63\,$\pm$\,0.55,\,\,$\Gamma$\,=\,1.76\,$\pm$\,0.17 and $\Gamma_{\text{comp}}$\,=\,1.67\,$\pm$\,0.19. The values for $kT$ and $\tau$ for these 144 selected points are $kT$\,=\,43\,$\pm$\,17 (keV) and $\tau$\,=\,1.5\,$\pm$\,0.5; their correlation is presented in Figure \ref{Figure04}.

\begin{figure}[ht]
\centerline{\includegraphics[width=0.9\columnwidth,keepaspectratio,angle=0]{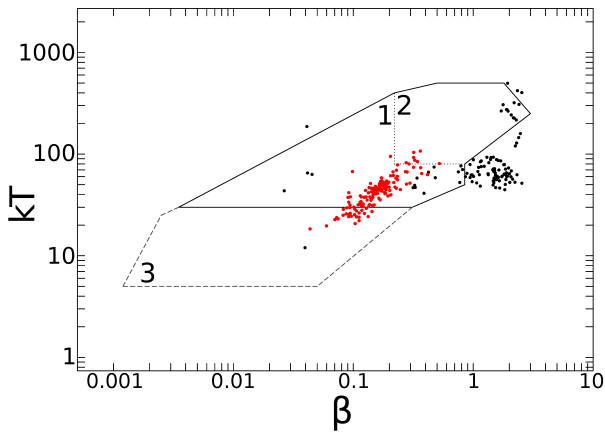}}
\caption{Diagram, based on and calculated from \citet{HuaTita95}, indicating the applicability zone of \texttt{comptt} and the location of our 250 spectra. Within the zone are 171 data points; drawn in red are the selected 144 whose $\Gamma$ and $\Gamma_{\text{comp}}$ were in good agreement.\label{Figure05}
        } 
\end{figure}

Based on the diagram presented in Figure 7 of {\citet{HuaTita95}}, we indicate, in Figure \ref{Figure05}, the position of all 250 spectra in the applicability zone aforementioned. The $\beta$ parameter, that depends only on $\tau$, was calculated for each spectra from Equation 3 of {\citet{HuaTita95}} for the non-spherical geometry case. 
In the diagram of Figure \ref{Figure05}, data points inside regions 1 and 2 correspond to characteristic spectra of optically thick and thin coronas, respectively; in region 3, the classification becomes less clear \citep{HuaTita95}. Points within any of these regions are inside the applicability zone of \texttt{comptt}. The 144 cases in which $\Gamma$ and $\Gamma_{\text{comp}}$ agreed within 20\% points are drawn in red.

\section{Discussion and final remarks}
We have presented the results from modelling a great number of hard X-ray (20--200\,keV) spectra of the black hole candidate \E. An (almost) immediate outcome of the analysis was that, thanks to the extension and quality of INTEGRAL database, we were able to produce what is perhaps the longest hard X-ray light curve of 1E, which, due to recent estimates of the distance to and the mass of the compact object, could be displayed here in the convenient units of $L$/$L_{\text{Edd}}$. 

It is worth mentioning that $\Gamma$ provided for the 283 spectra fitted by the \texttt{powerlaw} varied between $\sim$1.3--2.4, with a median value of 1.8\,$\pm$\,0.2. This range is compatible with a black hole in the low/hard state \citep{RemMc2006} and reaffirms 1E's permanency in this condition.

Regarding the corona, after employing a few selection criteria to avoid misusing the results of \texttt{comptt}, we have come to a total of 171 spectra. In 144 (roughly 85\%), the equation to compute $\Gamma_{\text{comp}}$ from the corona's physical parameters worked very well in ``reproducing'' the spectral shape observed (using as proxy the slope from a power-law fit). This suggests that the (hard) spectral variations of 1E may be explained by means of only $kT$ and $\tau$. Recalling that the approximation is valid only for the unsaturated comptonization regime, we can state that this is the predominant regime for the corona in 1E. Moreover, their position in the diagram of Figure \ref{Figure05} indicate a predominantly optically thick corona as well. A closer look at the spectra that deviated from a power-law and at the few whose indices did not match the ones calculated from $kT$ and $\tau$ is required for a more complete characterization and understanding of 1E's corona behavior.

It is discussed in \citet{Banerjee2020} that a $kT$\,$\times$\,$y$ diagram may be used to distinguish whether the compact object in a X-ray binary is a neutron star or a black hole. Considering the (likely) black hole  nature of 1E, we can assert that our results marginally corroborate the authors' claim that black holes occupy a slightly upper region in such diagram (see, e.g., their Figure 1).

At last, however we find that the spectral changes in \E\space are, most of the time, driven by variations in the corona, neither $\Gamma$ nor $kT$ and $\tau$ seem to be correlated with the 20--200\, keV flux observed. Two possible explanations for the long flux excursions (see, e.g Figure \ref{Figure01}) are \textit{i)} a geometrically stable corona that is eventually partially obscured or \textit{ii)} a corona whose size is in fact changing. By any means, our study does not allow us to make further assertions regarding this matter. 

\section*{Acknowledgements}

PES acknowledges FAPESP for financial support
under grant {\#}2017/13351-6. JL acknowledges Instituto Federal de S\~ao Paulo, IFSP/CAR,
{\#}2018/2.340.

\bibliographystyle{Wiley-ASNA}
\bibliography{TheBibFile}

\end{document}